\documentclass[useAMS,usenatbib]{mn2e}
\usepackage{graphicx}

\title[Gamma-ray emission from globular clusters]
  {Gamma-ray emission from globular clusters 2MS-GC01, IC 1257, NGC5904, NGC6656 and FSR1735}
\author[J. N. Zhou et al.]
  {J.~N.~Zhou$^{1,2}$, P.~F.~Zhang$^{1,3}$, X.~Y.~Huang$^{1,4}$\thanks{E-mail: xyhuang@pmo.ac.cn}, X.~Li$^{1,2}$, Y.~F.~Liang$^{1,2}$, L.~Fu$^{1}$,
  \newauthor
 J.~Z.~Yan$^{1}$, Q.~Z.~Liu$^{1}$\\
  $^{1}$ Key Laboratory of Dark Matter and Space Astronomy, Purple Mountain Observatory, Chinese Academy of Sciences, Nanjing 210008, China\\
  $^{2}$ University of Chinese Academy of Science, Yuquan Road 19, Beijing, 100049, China \\
  $^{3}$ Department of Physics, Yunnan University, Kunming, China\\
  $^{4}$ Physik-Department T30d, Technische Universit$\ddot{a}$t M$\ddot{u}$nchen, James-Franck-Stra$\ss$e, 85748 Garching, Germany\\}

\begin{document}

\label{firstpage}
\maketitle

\begin{abstract}
Globular clusters (GCs) are the oldest stellar system in the Galaxy, in which the millisecond pulsars are widely believed to be the only steady $\gamma$-ray emitters. So far 9 $\gamma$-ray GCs have been identified and a few candidates such as 2MS-GC01 and IC 1257 have been suggested. In this work, after analyzing the publicly-available {\it Fermi} LAT data we confirm the significant $\gamma$-ray emission from 2MS-GC01 and IC 1257 and report the discovery for $\gamma$-ray emission from NGC 5904 and NGC 6656 within their tidal radii. Also a strong evidence of  significant $\gamma$-ray emission is found from FSR 1735. From the observed $\gamma$-ray luminosities, the numbers of MSPs that are expected to be present in these GCs are estimated.
\end{abstract}


\section{Introduction}
Globular clusters (GCs) are spherical stellar systems in which stars are tightly bound by gravity. They are the oldest and also the most dense stellar systems in the Galaxy. On average the stellar density is $\sim$ 0.4 star per cubic parsec, increasing to 100 or 1000 stars per cubic parsec in their cores. The dynamical interactions with high stellar encounter rate are expected to form low-mass X-ray binaries (LMXBs). The formation rate per unit mass of LMXBs in GCs is known to be greater than in other places of the Galaxy \citep{katz,clark} and millisecond pulsars (MSPs) are generally believed to be the descendants of LMXBs \citep{alpar}. It is also widely accepted that MSPs contribute almost all $\gamma$-ray emissions of GCs. Before 2008 no GCs had been known as  $\gamma$-ray emitter. The situation changed dramatically after the successful launch of {\it Fermi} Gamma-ray Space Telescope \citep{Atwood2009} since the {\it Large Area Telescope} (LAT) onboard the {\it Fermi} satellite,
a pair-production telescope, can detect $\gamma$-rays at energies between $\sim$ 20 MeV and $>$ 300 GeV with unprecedent sensitivity. The discovery of $\gamma$-ray emission from GCs was made in 47 Tucanae (NGC 104; Abdo et al. 2009), in which at least 23 MSPs have been known by radio/X-ray observations.
As suggested in \citet{abdo2010}, the $\gamma$-ray emission from GCs is not due to single MSP but all the MSPs as a population.

More than 150 GCs have been identified in optical band \citep{harris}, but most of them are not significant $\gamma$-ray emitters.  
Nine 2FGL sources are doubtless associated with GCs, including 47 Tuc \citep{abdo2009}, Omega Cen (NGC 5139), NGC 6266, NGC 6388, Terzan 5, NGC 6440, NGC 6626, NGC 6652 \citep{abdo2010} and M 80 \citep[NGC~6093;][]{tam}. Moreover, there are a few candidates such as 2MS-GC01 and IC 1257 \citep{nolan2012}. The non-detection of significant $\gamma$-ray emission from most GCs is likely due to their low luminosities of the GeV emission and due to the strong background emission for the global clusters superposed in the Galactic disk.

In this work, we analyze the latest {\it Fermi} LAT data aiming to identify more $\gamma$-ray emitters of GCs. This work is structured as the following. In section 2 we present our data analysis. Our results are introduced in Section 3. Section 4 is our summary with some discussion.

\section{DATA ANALYSIS}

\subsection{Data preparation}
We use more than 6 years of {\it Fermi} LAT data (Pass 7 Reprocessed data; totally covering 2191 days from August 4th 2008) with energy ranges from 100 MeV to 100 GeV. The Fermi Science Tools {\tt v9r33p0} package is used to analyze the data, which is available from the Fermi Science Support Center\footnote{http://fermi.gsfc.nasa.gov/ssc/data/analysis/software/}. The {\tt P7REP\_SOURCE} (evclass=2) photons for point source analysis are selected in this work. To reduce the contribution from Earth albedo $\gamma$-rays, we only use events with zenith angles $\leq 100^{\circ}$. The {\tt P7REP\_SOURCE\_V15} instrument response functions (IRFs) is used.

\subsection{Analysis method}
We try an overall search of the GCs listed in the catalogue compiled by Harries (1996). In the analysis of each GC, photons from a 20$^{\circ}$ $\times$ 20$^{\circ}$ square region of interest (ROI) are selected, and binned into spatial pixels of 0.1$^{\circ}$ $\times$ 0.1$^{\circ}$. The first step is to apply standard binned maximum likelihood analysis using {\it gtlike}. We modelled the events with components of target source (with coordinates given in the catalogue) and backgrounds. The background components in this model are composed of all sources in the second {\it Fermi} LAT catalogue\footnote{http://fermi.gsfc.nasa.gov/ssc/data/access/lat/2yr\_catalog/} within the ROI and diffuse components, including the Galactic diffuse model ({\tt gll\_iem\_v05\_rev1.fit}) and isotropic background ({\tt iso\_source\_v05\_rev1.txt}). As usual, a single power law (PL) is used to model the point source. We also apply an exponential cutoff power law (PLE) for the $\gamma$-ray GC candidates, in which radio pulsars have been detected. For both models, the prefactor and the spectral index are set free. Additionally, the cut off energy is set free in PLE model.

We then choose the GCs with Test Statistic (TS) values greater than 25 for a further study. The TS is defined as TS = 2($\mathcal{L}1 - \mathcal{L}$0), where $\mathcal{L}$0 is the logarithmic maximum likelihood value for a model without the source (null hypothesis) and $\mathcal{L}$1 for a model with additional source at a specified location. For those $\gamma$-ray GC candidates with TS $\ge$ 25, we created  $5^{\circ} \times 5^{\circ}$ residual TS maps for them using the tool {\it gttsmap}. The TS maps aiming to identify weaker sources are created by moving a putative point source through a grid of locations on the sky and maximizing {\it -log(Likelihood)} at each grid point\footnote{http://fermi.gsfc.nasa.gov/ssc/data/analysis/scitools/likelihood\_t
utorial.html}. In taking this step, the target source that corresponds to the GC is removed from the model. All point source parameters are fixed while the diffuse components are set free. To reduce the contamination from background emissions due to point spread function (PSF) at lower energy band, different energy ranges for each GC are selected in TS map creation.

\begin{figure*}
\centering
	\includegraphics[scale=0.4]{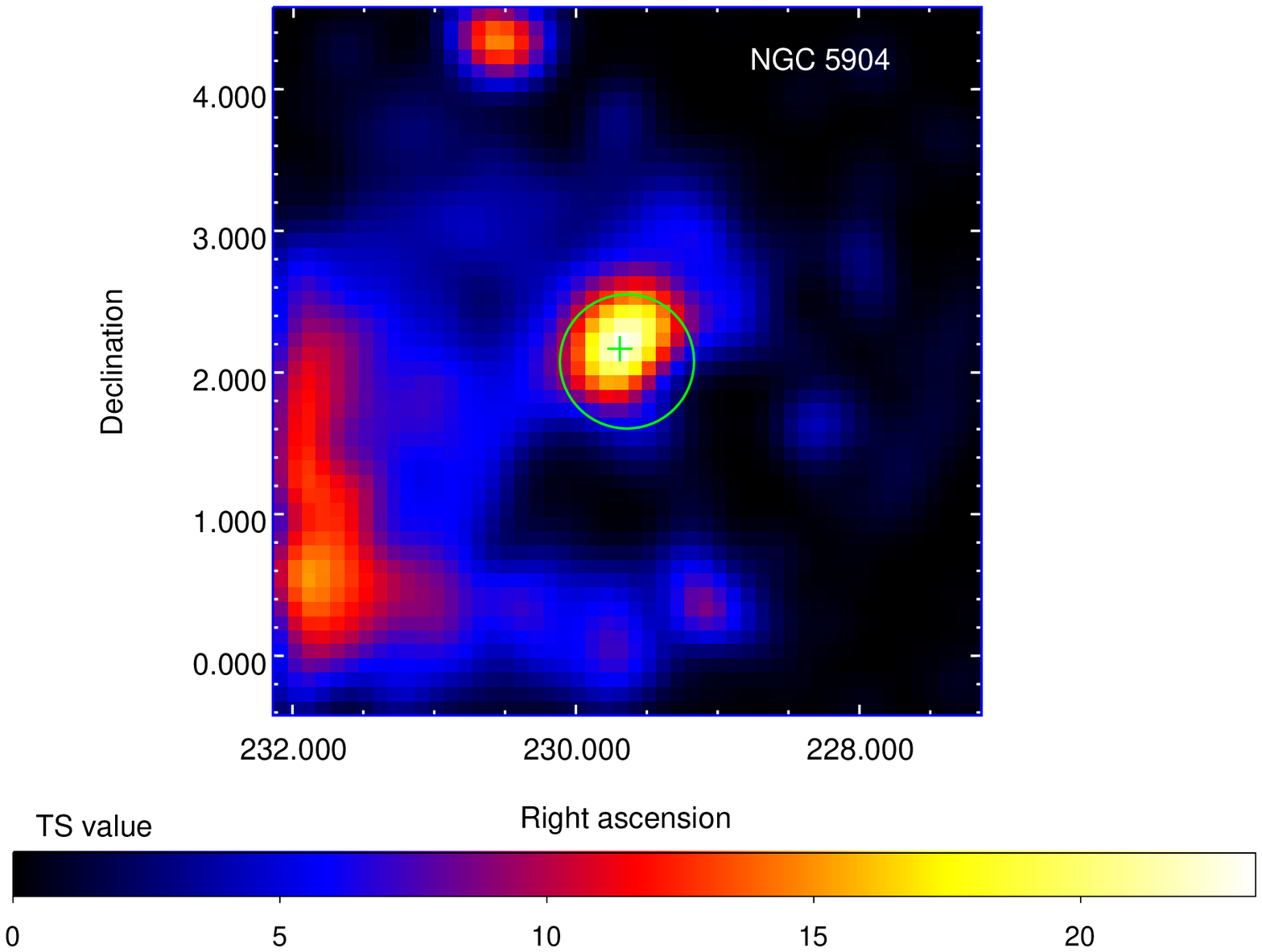}
	\includegraphics[scale=0.4]{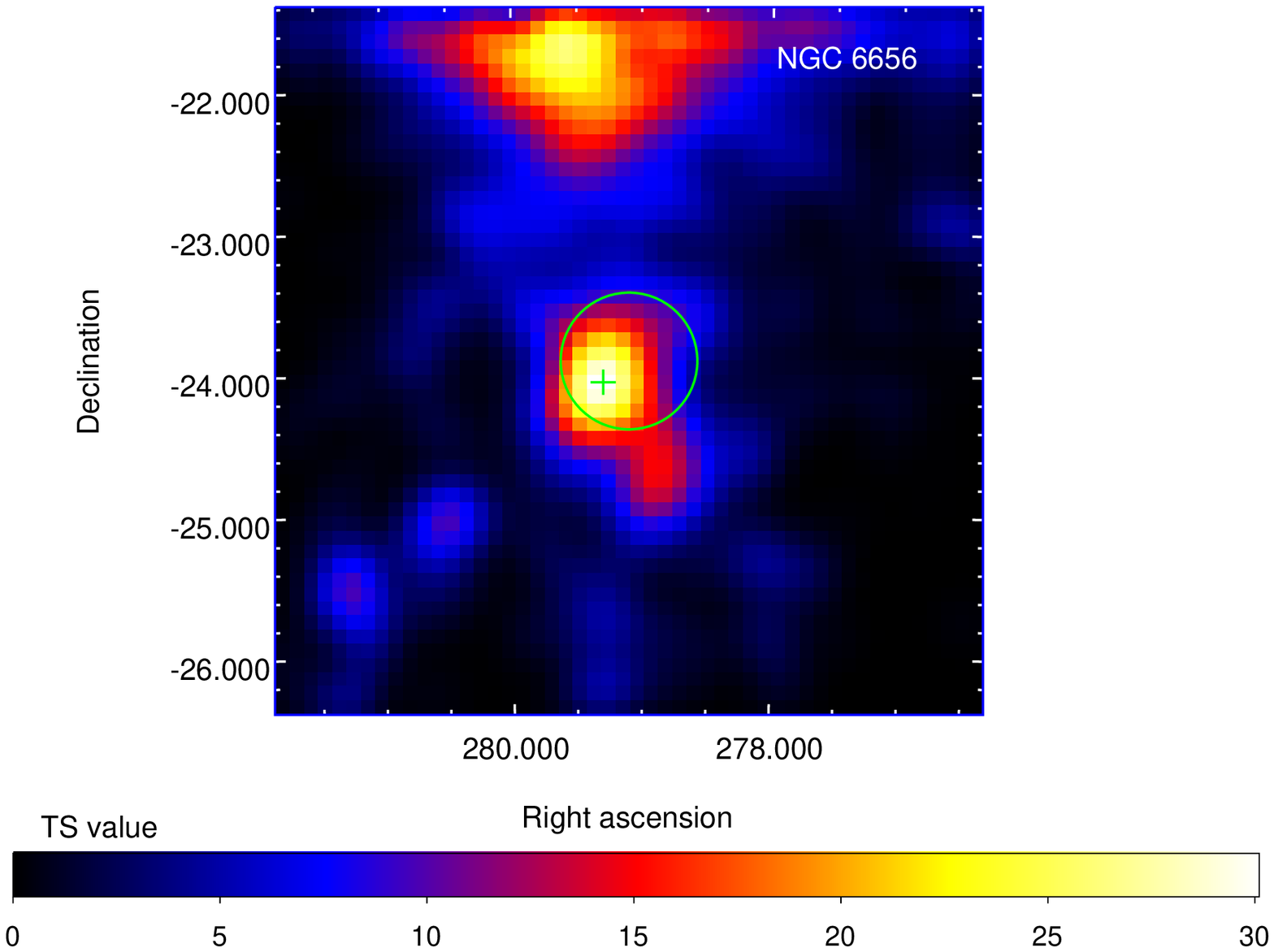}
	\includegraphics[scale=0.4]{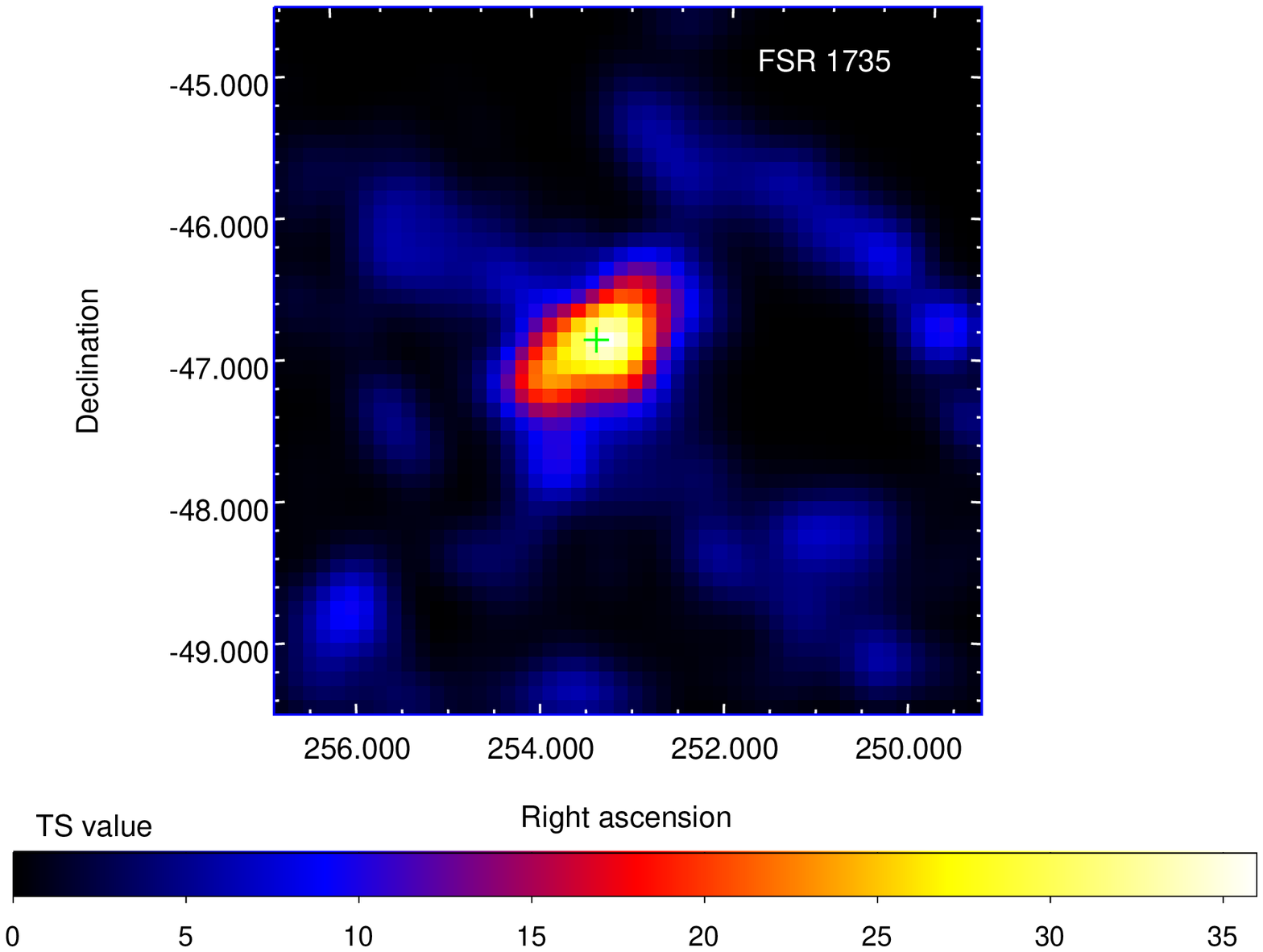}
		\caption{5$^{\circ}$ $\times$ 5$^{\circ}$ TS maps of the ROIs for NGC 5904, NGC 6656 and FSR 1735. The cross represent the best-fit centroid of the $\gamma$-ray emission from them. The circles represent the tidal radii \citep[][ 2003 version]{harris}. The tidal radius of FSR 1735 is unknown.}
	\label{3TSmaps}
\end{figure*}

From the residual TS maps, we find NGC 5904, NGC 6656 have evidences of $\gamma$-ray emission within their tidal radii. Meanwhile, there is a $\gamma$-ray emission evidence from FSR 1735 (see lower panel in Fig.~\ref{3TSmaps}), the separation between the best-fit position of emission and core of FSR 1735 is $15.6'$. Furthermore, \citet{nolan2012} reported other 2 associations with 2FGL sources in second catalogue: IC1257 (2FGL J1727.1-0704) and 2MS-GC01 (2FGL J1808.6-1950c). We also create TS maps for them. We then use {\it gtfindsrc} to search for the positions of these $\gamma$-ray candidates.

The significance from corresponding TS values follows the $\chi^{2}$ distribution with 2 degrees of freedom (PL model; 3 degrees for PLE model). To avoid fake signal with a number of search positions, similar estimation of trial factor is involved \citep{tam}. Totally 35 GCs with TS $>$ 25 are further analysed in this work, so $N_{GC} = 35$. The number of bins, $N_{bin} \sim 15$, is derived by dividing the averaged tidal radius area of all known GCs into 0.1$^{\circ}$ $\times$ 0.1$^{\circ}$ pixel, where the average tidal radius is $13'$. We get the post-trial significance using $(1 - {\it p})^{f} = 1 - {\it p'}$, where {\it p} is the pre-trial {\it p}-value, ${\it f} = N_{GC} \times N_{bin}$ is the trial factor, and ${\it p'}$ is the post-trial {\it p}-value.

Finally, we obtain the spectral energy distributions (SED) of these GCs. We divide the energy range from 200 MeV (100 MeV for 2MS-GC01) to 100 GeV into logarithmically equally spaced energy bins (base = 1.94). The flux is obtained by fitting all model components in each bin using {\it gtlike}. In the source model, the normalizations of Galactic diffuse component and isotropic component are left free.

\section{RESULTS}

\subsection{2MS-GC01}
\begin{figure*}
\centering
	\includegraphics[scale=0.45]{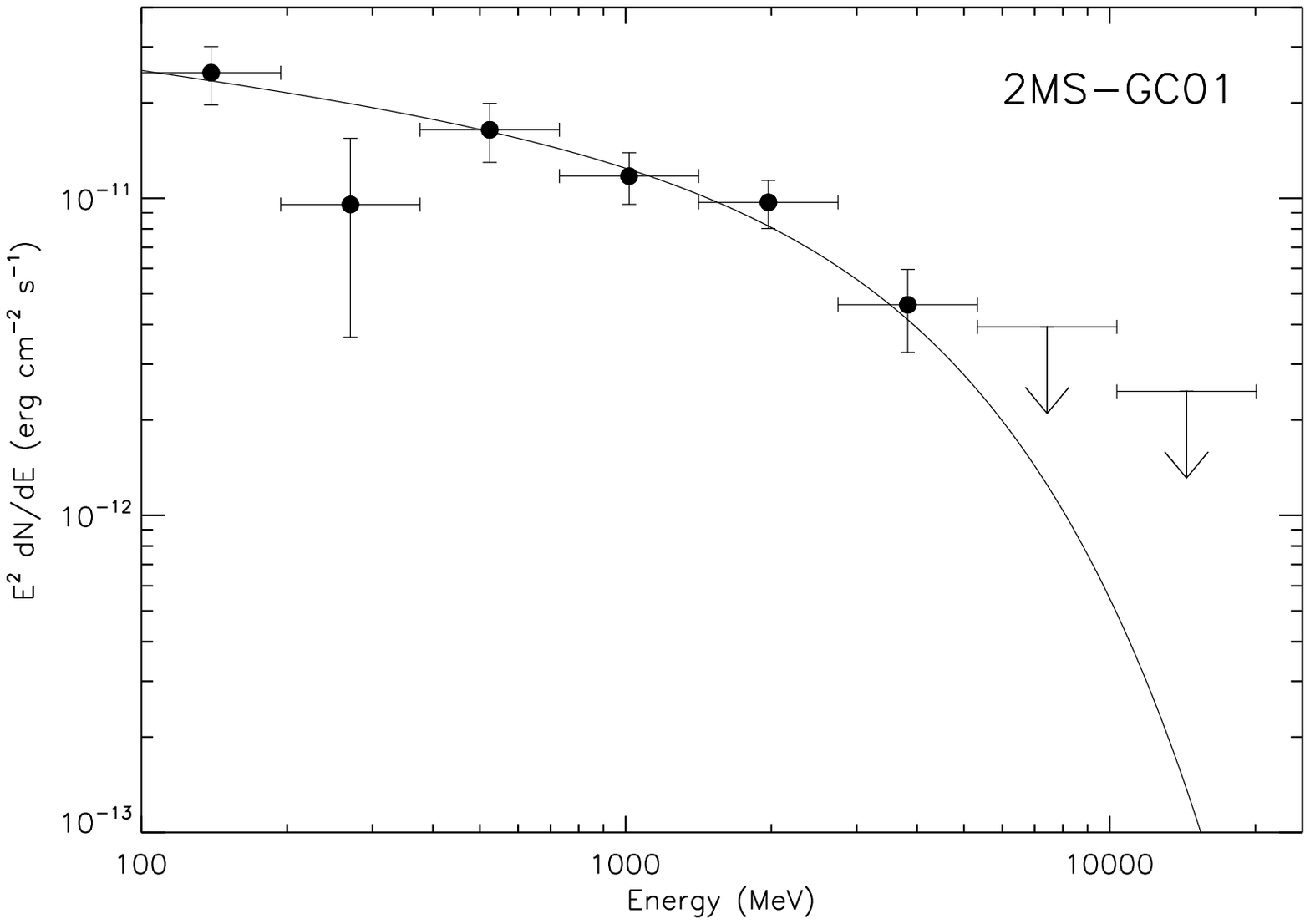}
	\includegraphics[scale=0.45]{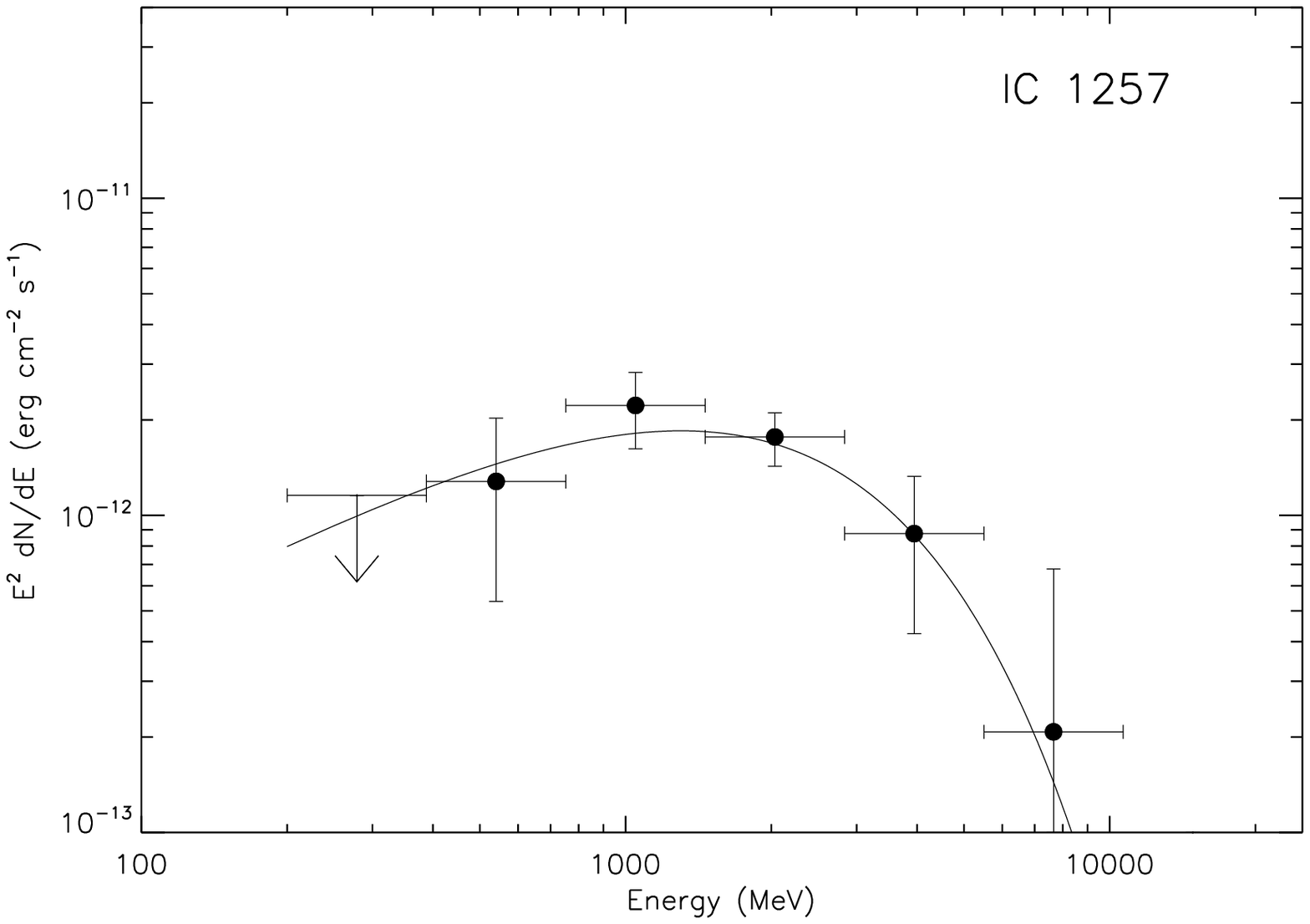}
	\includegraphics[scale=0.45]{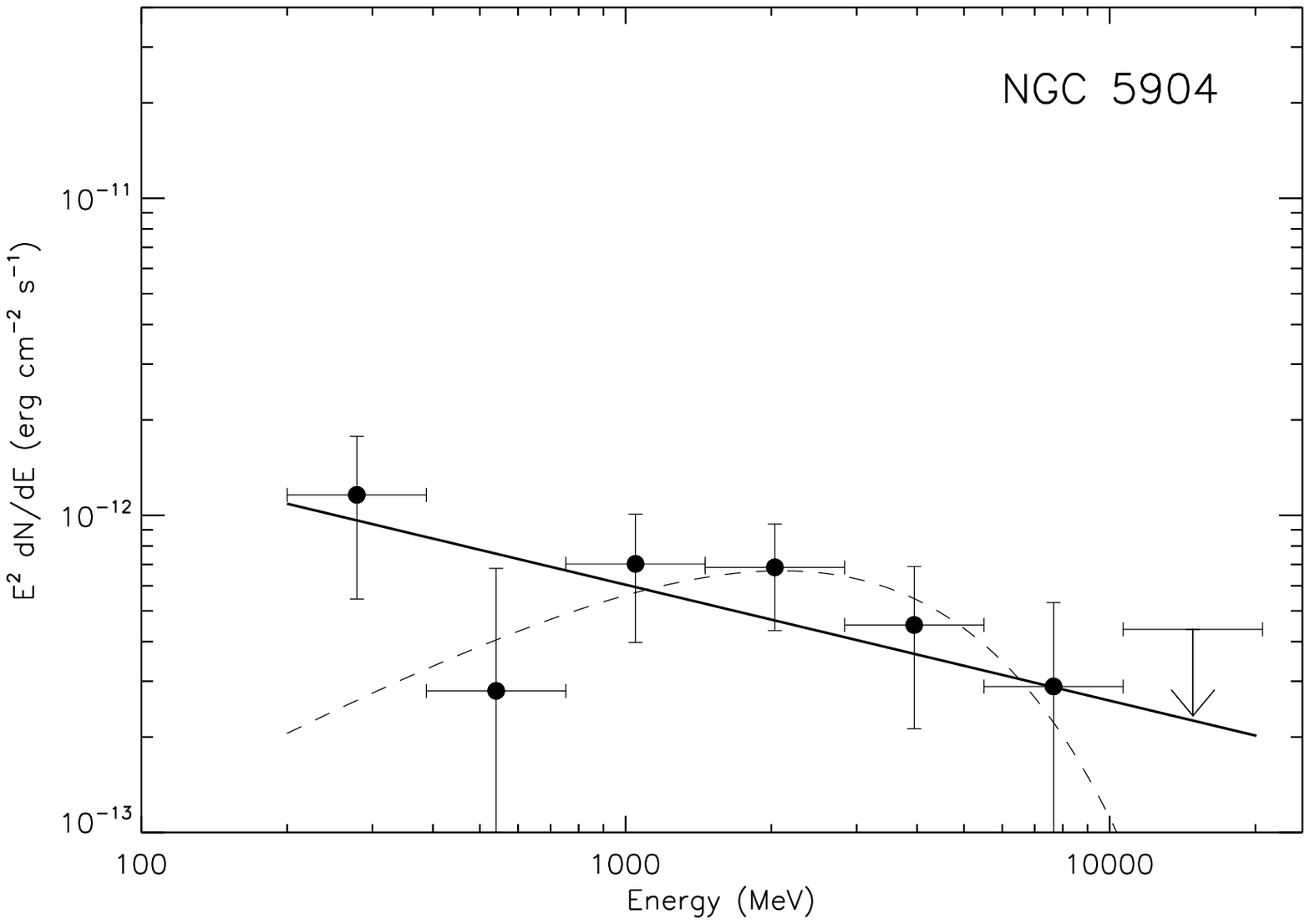}
	\includegraphics[scale=0.45]{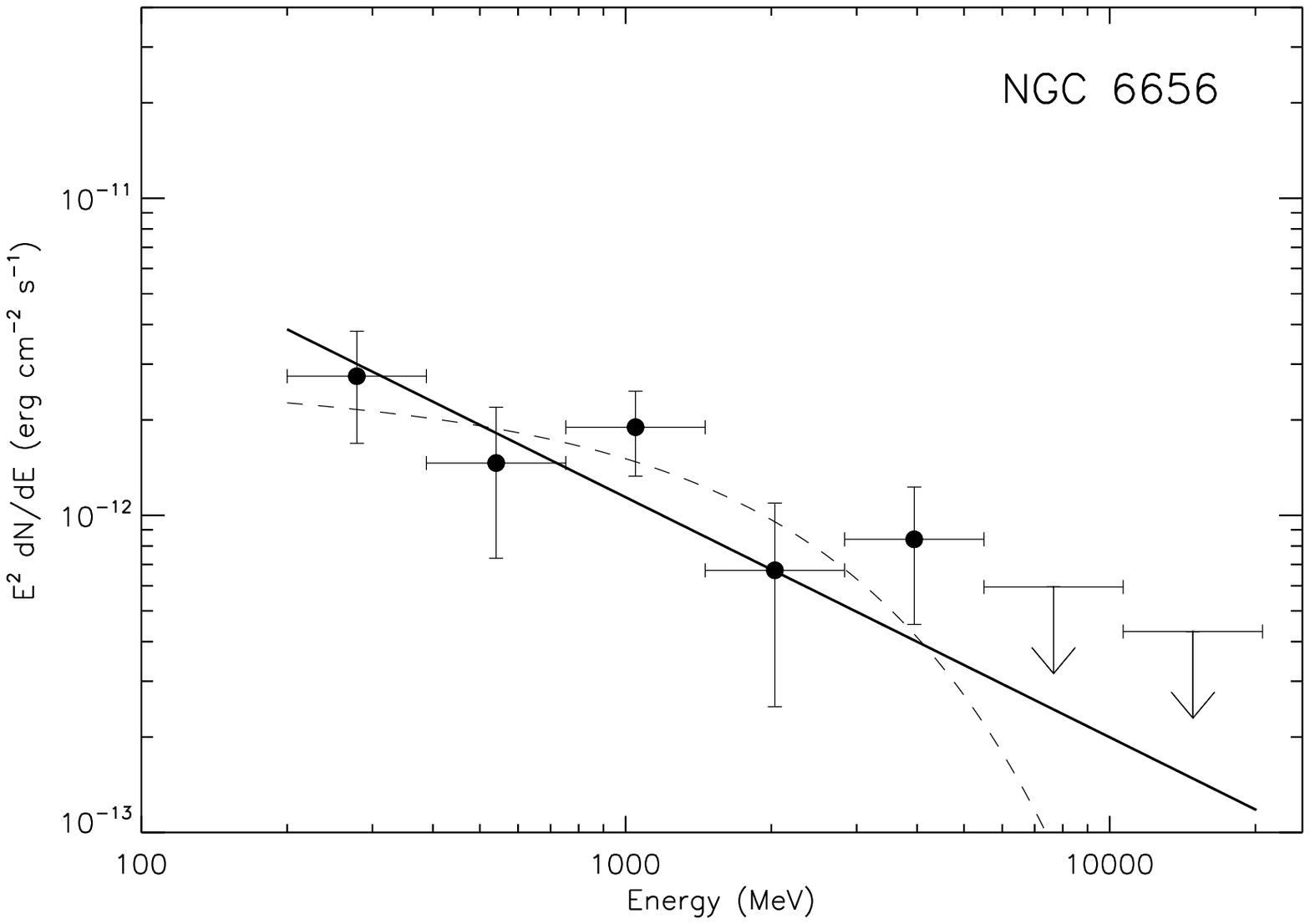}
	\includegraphics[scale=0.45]{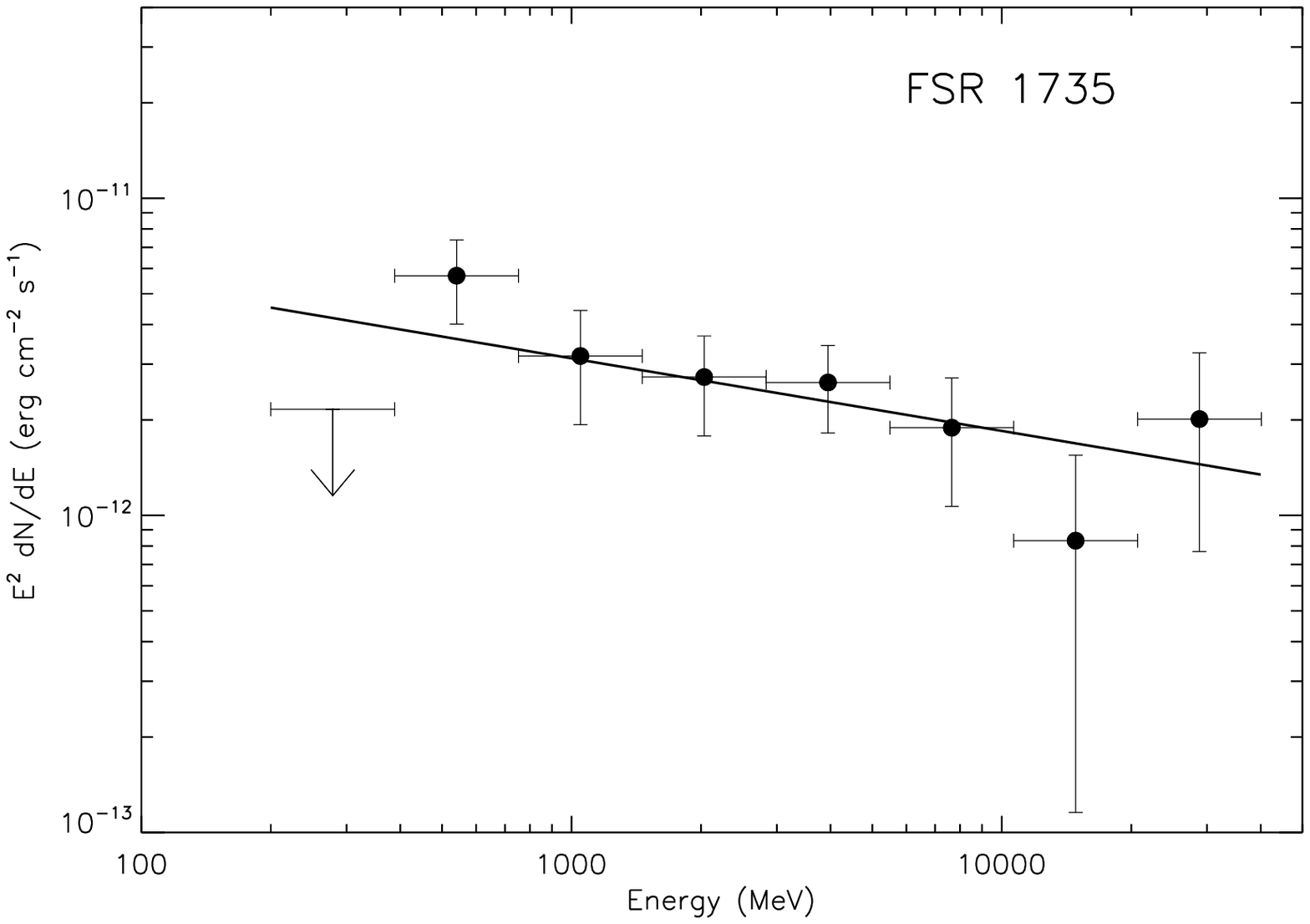}
		\caption{Spectral energy distributions of GCs. The solid lines indicate the bset fitted spectral model, while the dash lines indicate spectral fitting with a plausible model.}
	\label{sed}
\end{figure*}

\begin{figure*}
\centering
	\includegraphics[scale=0.4]{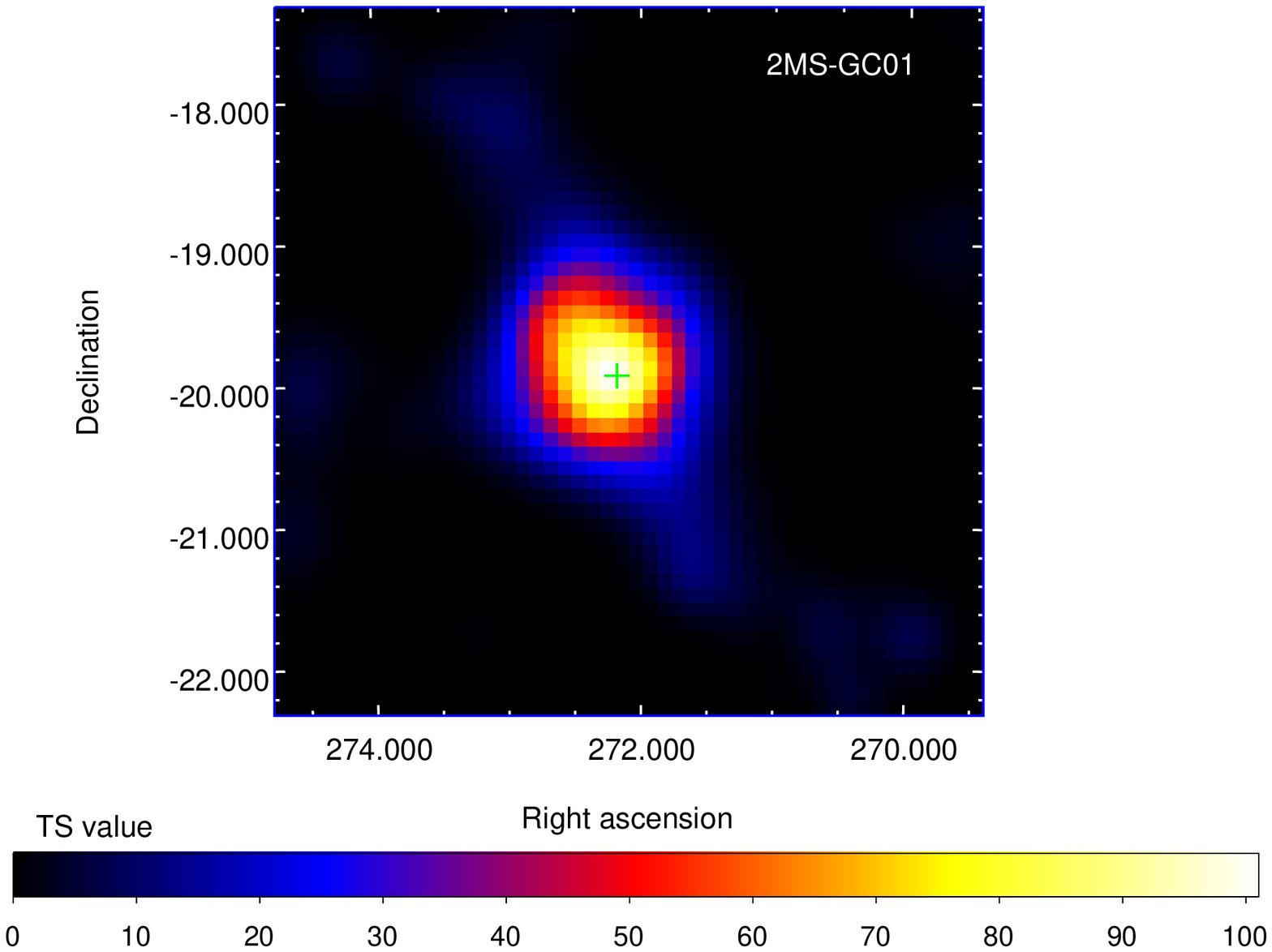}
	\includegraphics[scale=0.4]{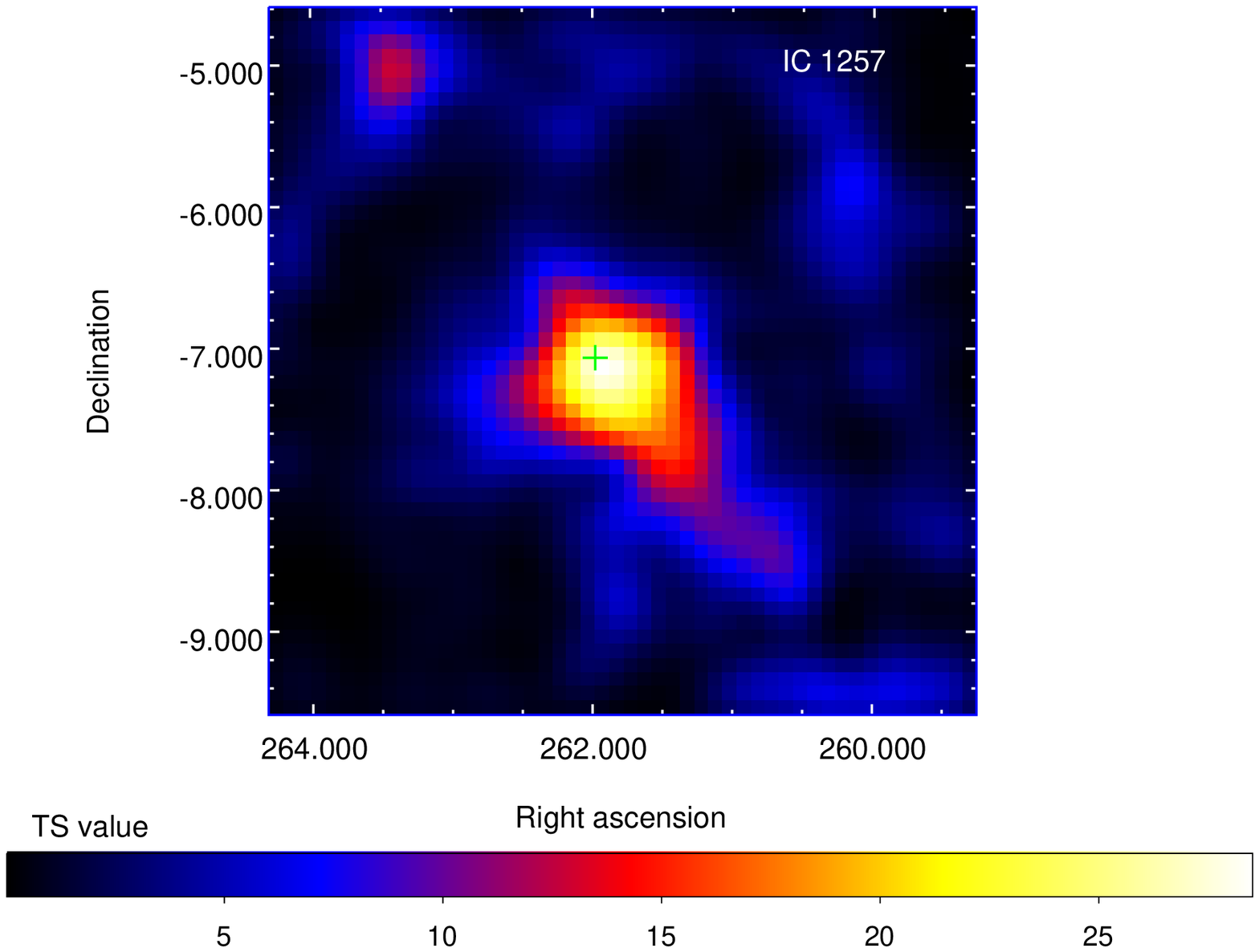}
		\caption{5$^{\circ}$ $\times$ 5$^{\circ}$ TS maps of the ROIs for 2MS-GC01 and IC 1257. Both tidal radii are unknown. The cross represent the best-fit centriod of the $\gamma$-ray emission from them.}
	\label{2TSmaps}
\end{figure*}

2MS-GC01 has formally been associated with 2FGL J1808.6-1950c \citep{nolan2012,cholis}. 
Our best-fit position is R.A. = 272.18, decl. = -19.93 (J2000), which is offset by $4.8'$ from the GC core (see the TS map, left panel of Fig.~\ref{2TSmaps}). We obtain an index of 2.2 $\pm$ 0.1 and a cutoff energy E$_{c}$=3.4 $\pm$ 0.2 GeV in a PLE spectral fitting. A TS value of 241, corresponding to a detection significance of 15.1$\sigma$ ($14.8\sigma$ post-trial) is given. If we just take into account the photons above 400 MeV (energy range selection for creating TS map), the detection significance is $\sim 10\sigma$. Fig.~\ref{sed} shows the spectrum of the $\gamma$-ray emission in E$^{2}$dN/dE. The integrated photon flux in the energy of 100 MeV$-$100 GeV is $F_{\rm 0.1-100 GeV} = (12.4 \pm 1.2) \times10^{-8}~{\rm cm}^{-2}~{\rm s}^{-1}$ and the integrated energy flux is $E_{\rm 0.1-100 GeV} = (5.7 \pm 0.5) \times 10^{-11}~{\rm erg}~{\rm cm}^{-2}~{\rm s}^{-1}$. At a distance of 3.6 kpc \citep[][ 2010]{harris}, the $\gamma$-ray luminosity is $L_{\rm 0.1-100 GeV} = (9.2 \pm 0.8) \times 10^{34}~{\rm erg~ s^{-1}}$.

\subsection{IC 1257}
IC 1257 has formally been associated with 2FGL J1727.1-0704 \citep{nolan2012,cholis}. Our best-fit position is R.A. = 261.98, decl. = -7.07 (J2000), which is offset by $10.6'$ from the cluster core (see the right panel in Fig.~\ref{2TSmaps}). Fitting the spectrum with a PLE model we have an index of $1.2 \pm 0.7$ with a cutoff energy E$_{c}$ = $1.6 \pm 0.9 $ GeV. We obtained a TS value of 35 and the corresponding detection significance of $5.2\sigma$ ($4.0\sigma$ post-trial). We have TS $= 30$ if instead just the photons above 400 MeV have been taken into account, corresponding to a significance of $4.7\sigma$ ($3.4\sigma$ post-trial). It is also the energy range selection for TS map creating. The integrated photon flux between 100 MeV and 100 GeV is $F_{\rm 0.1-100 GeV} = (5.7 \pm 1.0)  \times  10^{-8} ~{\rm cm}^{-2} {\rm s}^{-1}$, while the integrated energy flux is $E_{\rm 0.1-100 GeV} = (5.0 \pm 0.9)\times 10^{-12} ~{\rm erg~ cm^{-2}~s^{-1}}$. The luminosity is $L_{\rm 0.1-100 GeV} = (3.7 \pm 0.7) \times 10^{35}~{\rm erg~ s^{-1}}$ at a distance of 25 kpc. Such a luminosity is rather high in comparison with other $\gamma$-ray GCs, which is actually the largest among all detected $\gamma$-ray emitting GCs by now. 

\subsection{NGC 5904}
In this GC five radio pulsars\footnote{http://www.naic.edu/~pfreire/GCpsr.html} have been detected \citep{anderson,hessels} but no significant $\gamma$-ray emission has been reported. In our analysis we find some evidence for $\gamma$-ray emission from NGC 5904. With the photons in the energy range of $0.1-100$ GeV, the TS value under PL model is 26, corresponding to a significance of $4.6\sigma$ (see the upper left panel in Fig.~\ref{3TSmaps}) and the post-trial significance is $3.2\sigma$. To minimize the influence of strong background emission at low energies, we choose the photons at energies of $\geq500$ MeV (for these photons the PSF is much smaller than that at energies of $\sim 100$ MeV and the background pollution can be effectively suppressed) to derive TS map, which gives a TS value of 21 ($\sim$4.0$\sigma$; post-trial $2.4\sigma$). Our best-fit position is R.A. = 229.69, decl. = 2.17 (J2000), $5.9'$ from the core position of NGC 5904, well within the tidal radius of NGC 5904 \citep[][ the 2003 version]{harris}. The SED can be well fit by a single power law with an index of 2.3 $\pm$ 0.2. The integrated photon flux is $F_{\rm 0.1-100 GeV}$ = (6.2 $\pm$ 1.4) $\times$ 10$^{-9}$ cm$^{-2}$ s$^{-1}$ and the integrated energy flux is $E_{\rm 0.1-100 GeV}$ = (3.5 $\pm$ 0.8) $\times$ 10$^{-12}$ erg cm$^{-2}$ s$^{-1}$. When deriving TS map, an additional source (R.A. = 230.54, decl. = 4.32) is added to the model, of which the spectrum is set as a single power law and to the ROI center with a separation of 2.4$^{\circ}$. At a distance of 7.5 kpc, the $\gamma$-ray emission luminosity is $L_{\rm 0.1-100 GeV}$ = (2.4 $\pm$ 0.5) $\times$ 10$^{34}$ erg s$^{-1}$. We also use a PLE model to fit the spectrum, where the photon index $\Gamma = 1.2 \pm 1.3$ and the cutoff energy is (2.6 $\pm$ 2.8) GeV. In this model, the significance keeps on the same level, 3.1$\sigma$ post-trial. No significance of the PLE model over the PL model, so  we consider the PL model as a better model. 

\subsection{FSR 1735}
The $\gamma$-ray emission from FSR 1735 is found with a TS value of 55 and the corresponding significance is $\sim 7.0\sigma$ ($6.2\sigma$ post-trial). Again, we only take into account the photons above 400 MeV to create TS map (see lower panel in Fig.~\ref{3TSmaps}), the value of TS is $39$ and the corresponding significance is $5.8\sigma$ ($4.8\sigma$ post-trial). Our best-fit position is R.A. $= 253.37$, decl. $= -46.91$ (J2000). Though the tidal radius of FSR 1735 is unknown, the bset-fit $\gamma$-ray position found in our analysis just has an offset $\sim 15.6'$ from the core of FSR 1735. Using a single power law model, we obtained a photon index $\Gamma = 2.2 \pm 0.1$. The integrated photon flux is $F_{\rm 0.1-100 GeV} = (2.7 \pm 0.4) \times 10^{-8} ~{\rm cm}^{-2} {\rm s}^{-1}$ and the integrated energy flux is $F_{0.1-100 GeV} = (1.8 \pm 0.3) \times 10^{-11} ~{\rm erg}~{\rm cm}^{-2}~{\rm s}^{-1}$. At a distance of 9.8 kpc, the $\gamma$-ray luminosity is $L_{\rm 0.1-100 GeV} = (2.1 \pm 0.3) \times 10^{35}~{\rm erg}~{\rm s}^{-1}$. 

\subsection{NGC 6656}
Two radio pulsars\footnote{http://www.naic.edu/~pfreire/GCpsr.html} have been discovered \citep{lynch} in this GC. It also has no association in the second catalogue of LAT. Our reduction on this region show an evidence of $\gamma$-ray emission. The best-fit position (R.A.=279.30 decl.=-24.05; J2000) is well within the tidal radius circle of NGC 6656. Using a single power law model (3.4$\sigma$ post-trial), we obtained a photon index of $\Gamma$ = 2.7 $\pm$ 0.1, and 0.1-100 GeV photon and energy fluxes of (2.3 $\pm$ 0.6) $\times$ 10$^{-8}$ cm$^{-2}$ s$^{-1}$ and (8.6 $\pm$ 1.9) $\times$ 10$^{-12}$ erg cm$^{-2}$ s$^{-1}$, respectively. The $\gamma$-ray luminosity is $L_{\rm 0.1-100 GeV} = (1.1 \pm 0.2) \times 10^{34}~{\rm erg}~{\rm s}^{-1}$ at a distance of 3.2 kpc. The spectrum also can be fitted with a PLE model, where the photon index $\Gamma$ = 2.0 $\pm$ 0.5 and cutoff energy $E_{c} = (2.4 \pm 1.8)$ GeV. Due to the significance under PLE model is also 3.4$\sigma$ post-trial, so we consider PL model as the better one.

\section{Summary}

In this work we have analyzed the {\it Fermi} LAT data of more than 150 GCs and detect or find some evidence of $\gamma$-ray emission from 3 GCs, i.e. FSR 1735, NGC 5904 and NGC 6656. Another two (2MS-GC01 and IC 1257) have been reported as $\gamma$-ray emission candidates in \citet{nolan2012}, of which the associations of significant $\gamma$-ray emission have been confirmed here. Among the three new detected candidates, the evidence of $\gamma$-ray emission from FSR 1735 is very significant (the post-trial significance is at a confidence level of $>5\sigma$). Though the tidal radius of such a GC is unknown, the $\gamma$-ray source found in our analysis just has an offset $\sim 15.6'$ from the core of FSR 1735, which is small enough to allow us to suggest an association with these two sources. The $\gamma$-ray emission signals from the other two GCs are weaker (the confidence level is above $3\sigma$ but smaller than $5\sigma$) and more data is needed to firmly establish their $\gamma$-ray emitter nature. 
{\it Fermi} satellite continues to collect the $\gamma$-ray data and more $\gamma$-ray GCs are expected to be identified. Nevertheless for the GCs within the Galactic plane the prospect is not very promising due to the strong background that can minimize the $\gamma$-ray signal effectively. More GCs may be detected in the future with a high angular resolution telescope dedicated to the sub-TeV (from $\sim 10$ MeV to $\sim 10$ GeV) $\gamma$-ray photon detection. For example, the proposed mission ``PAir-productioN Gamma-ray Unit" aims to have an angular resolution a factor of $\geq 5$ better than the currently operating {\it Fermi} Gamma-ray Space Telescope in the sub-GeV range (Wu et al. 2014). The high quality $\gamma$-ray maps will significantly
improve the identification of ``point-like" sources from extended and complicated diffuse $\gamma$-ray
background, which is thus very suitable to identify more $\gamma$-ray GCs.  

As already mentioned in Section 3, no pulsars have been reliably identified in these GCs except NGC 5904 and NGC 6656. The detection of $\gamma$-ray emission in turn provides us the chance to estimate the number of millisecond pulsars that are believed to be the unique stable $\gamma$-ray emitters in GCs. Our approach is following Abdo et al. (2010) and the number of MSPs in each GC is estimated using their Eq.(1).
Where $\langle$$\dot{E}$$\rangle$ = (1.8 $\pm$ 0.7)$\times$10$^{34}$ erg s$^{-1}$ is adopted as average spin-down power of pulsar, and the $\gamma$-ray luminosity conversion efficiency is set to 0.08. The results are shown in Table~\ref{para_spec}. For NGC 5904 and NGC 6656, the estimated numbers of MSPs are more than that found currently, implying that in the future more powerful radio telescopes (or arrays) have very promising detection prospects as long as these MSPs also radiate non-ignorable amount of energy in radio bands.

\section*{ACKNOWLEDGEMENTS}

We acknowledge the use of data from Fermi Science Support Center (FSSC) at NASA's Goddard Space Flight Center. JNZ thanks Dr. Yizhong Fan for help in improving the presentation. This work is supported in part by 973 Program of China under grant 2013CB837000, the National Natural Science Foundation of China under grants 11273064, 11443004;  the Strategic Priority Research Program "The Emergence of Cosmological Structures" of the Chinese Academy of Sciences, Grant No. XDB09000000., and by China Postdoctoral Science Foundation under grant 2014M551680.

\clearpage
\begin{table*}
\begin{center}
\caption{}
\begin{tabular}{ccrrrccrc}
\hline\hline
GC & \multicolumn{4}{c}{Position$^{(1)}$} & Tidal$^{(2)}$ & \multicolumn{2}{c}{Position$^{(3)}$} & Offset  \\
Name & R.A. & Decl. & {\it l.} & {\it b.} & radius($'$) & R.A. & Decl. & ($'$)   \\
\hline
 NGC 5904 & 229.64 & 2.08    & 3.86 &  46.80 & 28.40 & 229.69 & 2.17  & 5.9	\\
 NGC 6656 & 279.10 & -23.90  & 9.89 &  -7.55 & 28.97 & 279.30 &-24.05 & 14.2   \\
 FSR 1735 & 253.04 & -47.06  &339.18& -1.85  & ...   & 253.37 &-46.91 & 15.6	\\
 2MS-GC01 & 272.09 & -19.83  &10.47 & 0.10   & ...   & 272.18 &-19.93 & 4.8    \\
 IC 1257  & 261.79 &  -7.09  &16.53 & 15.14  & ...   & 261.98 &-7.07  & 10.6	\\
\hline
\end{tabular}
\label{para_pos}
\end{center}
{\bf Notes.}
(1) Coordinates derived from catalogue of \citet[][ 2003 version]{harris} and corresponding Galactic coordinates, (2) Tidal radius of FSR 1735, 2MS-GC01 and IC 1257 are not given, (3) Our best-fit position of $\gamma$-ray emission from GCs.
\end{table*}

\begin{table*}
\begin{center}
\caption{Parameters of $\gamma$-ray from GCs.}
\begin{tabular}{ccccccccccc}
\hline\hline
GC  & Spectral  & TS & Significance & Photon & Cutoff& photon$^{(1)}$  & Energy$^{(2)}$  & Distance$^{(3)}$ & L$^{(4)}$ & N$_{MSP}$ \\
Name & model &  & (post-trial;$\sigma$) & index & (GeV) & flux  & flux &  (kpc)   &    & \\
\hline

 NGC 5904 & PL &  26  & 3.2 & 2.3 $\pm$ 0.2&...& 6.2 $\pm$ 1.4& 3.5 $\pm$ 0.8 & 7.5 & 2.4 $\pm$ 0.5    & 16 $\pm$ 4   \\
          & PLE&  28  & 3.1 & 1.2 $\pm$ 1.3&2.6 $\pm$ 2.8& 1.6 $\pm$ 0.4 & 1.9 $\pm$ 0.4 & 7.5 & 1.3 $\pm$ 0.3 & 9 $\pm$ 2  \\
 NGC 6656 & PL &  27  & 3.4 & 2.7 $\pm$ 0.1&...& 23.2 $\pm$ 5.9& 8.6 $\pm$ 1.9 & 3.2 & 1.1 $\pm$ 0.2   & 7 $\pm$ 2   \\	
          & PLE&  30  & 3.4 & 2.0 $\pm$ 0.5&2.4 $\pm$ 1.8& 13.1 $\pm$ 4.1 & 6.3 $\pm$ 1.4 & 3.2 & 0.8 $\pm$ 0.2 & 5 $\pm$ 1  \\
 FSR 1735 & PL &  55  & 6.2 & 2.2 $\pm$ 0.1&...& 26.9 $\pm$ 4.3& 18.4 $\pm$ 2.9 & 9.8 & 21.1 $\pm$ 3.3 & 147 $\pm$ 23  \\
 2MS-GC01 & PLE&  241 & 14.8& 2.2 $\pm$ 0.1&3.4 $\pm$ 0.2& 124.7 $\pm$ 11.8& 56.8 $\pm$ 5.0 &3.6& 8.8 $\pm$ 0.8& 61 $\pm$ 5   \\
IC 1257   & PLE&  35  & 4.0  & 1.2 $\pm$ 0.7&1.6 $\pm$ 0.9& 5.7 $\pm$ 1.0 & 5.0 $\pm$ 0.9 & 25.0 & 37.4 $\pm$ 6.7& 260 $\pm$ 47 \\
\hline
\end{tabular}
\label{para_spec}
\end{center}
{\bf Notes.}
(1) Integrated 0.1--100~GeV photon flux in unit of 10$^{-9}$ cm$^{-2}$ s$^{-1}$, (2) Integrated 0.1--100~GeV energy flux in unit of 10$^{-12}$ erg cm$^{-2}$ s$^{-1}$, (3) Distance from \citet[][ 2003 version]{harris}, (4) 0.1--100~GeV luminosity in unit of 10$^{34}$~erg~s$^{-1}$.
\end{table*}

\clearpage
%

\label{lastpage}
\end{document}